\documentclass[aps,pre]{revtex4}
\usepackage{mathrsfs}
\usepackage{amssymb}
\usepackage{graphicx}
\usepackage{amsmath}
\usepackage{dcolumn}% Align table columns on decimal point
\usepackage{bm}% bold math

\usepackage[utf8]{inputenc} % allow utf-8 input
\usepackage[T1]{fontenc}    % use 8-bit T1 fonts
\usepackage{hyperref}       % hyperlinks
\usepackage{url}            % simple URL typesetting
\usepackage{booktabs}       % professional-quality tables
\usepackage{amsfonts}       % blackboard math symbols
\usepackage{nicefrac}       % compact symbols for 1/2, etc.
\usepackage{microtype}      % microtypography
\usepackage{bbm}
\usepackage{subfigure}
\newcommand{\A}{\mathbb{A}}
\newcommand{\D}{\mathbb{D}}
\newcommand{\mydelta}[1]{\delta\left(#1\right)}
\newcommand{\id}{\mathbb{I}}
\newcommand{\bra}[1]{\left(#1\right)}
\newcommand{\brb}[1]{\left[#1\right]}

\newcommand{\ind}{\mathbbm{1}}

\newcommand{\ein}{\epsilon_{\textrm{in}}}
\newcommand{\eout}{\epsilon_{\textrm{out}}}

\newtheorem{theorem}{Theorem}

\newtheorem{lemma}[theorem]{Lemma}

\newtheorem{definition}[theorem]{Definition}

\begin{document}

\title{Spectral estimation of the percolation transition in clustered networks}
\author{ Pan Zhang}
\email{panzhang@itp.ac.cn}
\affiliation{
CAS key Laboratory of Theoretical Physics, Institute of Theoretical Physics, Chinese Academy of Sciences, Beijing 100190, China.
}

\begin{abstract}
	There have been several spectral bounds for the percolation 
	transition in networks, using spectrum of matrices associated with
	the network such as the adjacency matrix and the non-backtracking matrix.
	However they are far from being tight when the network is sparse and displays
	clustering or transitivity, which is represented by existence of short loops e.g. triangles.
	In this work, for the bond percolation, we first propose a message passing algorithm for calculating 
	size of percolating clusters considering effects of triangles, 
	then relate the percolation transition to the leading eigenvalue of 
	a matrix that we name the \textit{triangle-non-backtracking matrix}, 
	by analyzing stability 
	of the message passing equations. 
	We establish that our method gives a tighter lower-bound to the 
	bond percolation transition than
	previous spectral bounds, and it becomes exact for an infinite 
	network with no loops longer than $3$.
%	Our approach is also numerically evaluated on both synthetic and 
%	real-world networks.
	We evaluate numerically our methods on synthetic and real-world
	networks, and discuss further generalizations of our approach 
	to include higher-order sub-structures.
\end{abstract}


\maketitle

\section{Introduction}
%Percolation on networks is a well-known process that has been used 
%	as a model of many processes such as epidemic spreading and resilience
%	of system. 
Percolation in networks is a well-known process that has been 
studied extensively in many fields of science. It found
numerous applications in physics, networks, material science and social science \cite{Newman2002spread,Cohen2000resilience,PhysRevLett.66.169,Bondt1992,PhysRevLett.71.2741,Albert2000error,RevModPhys.45.574,PhysRevB.4.2612,PhysRevLett.39.1222}.
In this work we study the bond percolation on an arbitrary undirected network, where each edge is open with
probability $p$ and close with probability $1-p$. In the $n\to\infty$
limit, depending on $p$, in the 
network there could be one giant cluster that contains a finite fraction 
of nodes in addition to many small clusters with vanishing
fraction of nodes, or no giant cluster at all. The transition value $p^*$
determines the smallest value of $p$ that with high probability 
the giant cluster emerges.

There have been lots of analytical studies of percolation transition on
synthetic networks, such as 
random graphs \cite{Cohen2000resilience,PhysRevE.65.056109,PhysRevE.78.051105,Callaway2000},
generalization of random graphs with triangles and cliques \cite{Gleeson2009bond,Gleeson2010how,miller2009percolation,Serrano2006prl,Serrano2006clustering} etc.
However most of those studies are specific for ensemble of networks in the
$n\to\infty$ limit, rather than a given (synthetic or real-world) network.
Recently there are studies on approximating the percolation transition
for a given real-world network using spectral properties of matrices
that associated with the network. Here we use ``approximate'' because 
the percolation transition is legally defined only for infinite networks, 
where an infinite giant cluster could appear.
On finite networks, one usually identifies the percolation transition by the point where the second-largest cluster has the greatest size, which converges to the percolation threshold when system size goes to infinity.

This can be done by running numerous direct simulations to compute the size of the largest connected component on each realization of random occupations of edges, then take the ensemble average. However this is time-consuming. 
In \cite{bollobas2010percolation} it is proposed to use inverse 
spectral radius of the network, that is inverse of the leading 
eigenvalue of the network's adjacency matrix $A$, as an estimate of 
the percolation transition. This estimate is accurate when
the network is dense, but heavily underestimates the percolation transition
on sparse networks.
In recent work \cite{Hamilton14,karrer2014percolation} the authors proposed to approximate
percolation transition on sparse networks using inverse of 
leading eigenvalue of
the non-backtracking matrix \cite{hashimoto1989zeta,Krzakala2013,Zhang2015}, which
is defined on directed edges of the network. In addition to computational speed, another motivation of using spectral methods for the percolation is that it provides more information on the percolation than the size of clusters in a compact way, which can be used to organize efficient algorithms for applications based on percolation, such as the network dismantling problem~\cite{Morone2015,Zdeborova2016}.

The idea behind the use of the non-backtracking matrix comes from 
the linearization of belief propagation (BP) \cite{Yedidia2001}
equations around a factorized
fixed point. Since the belief propagation assumes the conditional independence, it is exact when effects of loops in the network can be neglected, and is a good approximation in real-world networks. Authors in \cite{Hamilton14,karrer2014percolation} have shown
that the estimation of the percolation transition
using the non-backtracking matrix is a lower-bound to the true percolation
transition on an infinite undirected network, and is exact when
the network is an infinite tree. In \cite{karrer2014percolation} by
comparing the estimate to the direct simulation of percolation processes
on real-world networks, the authors also showed that the estimate is a better
approximation than the inverse for spectral radius for real-world networks,
since most of the real-world networks are sparse.

We note here that in addition to sparsity, another characteristic of real-world networks is
clustering, or transitivity, which is represented by existence of 
short loops like triangles in the network. However the estimate of percolation transition using
the non-backtracking matrix assumes that the network is locally tree like,
thus ignores effects of short loops. In this work we address the problem of how to 
incorporate the loops, especially, triangles, in estimating the percolation transition. 

\begin{figure}[h]
  \centering
\subfigure[]{
\includegraphics[width=0.4\columnwidth]{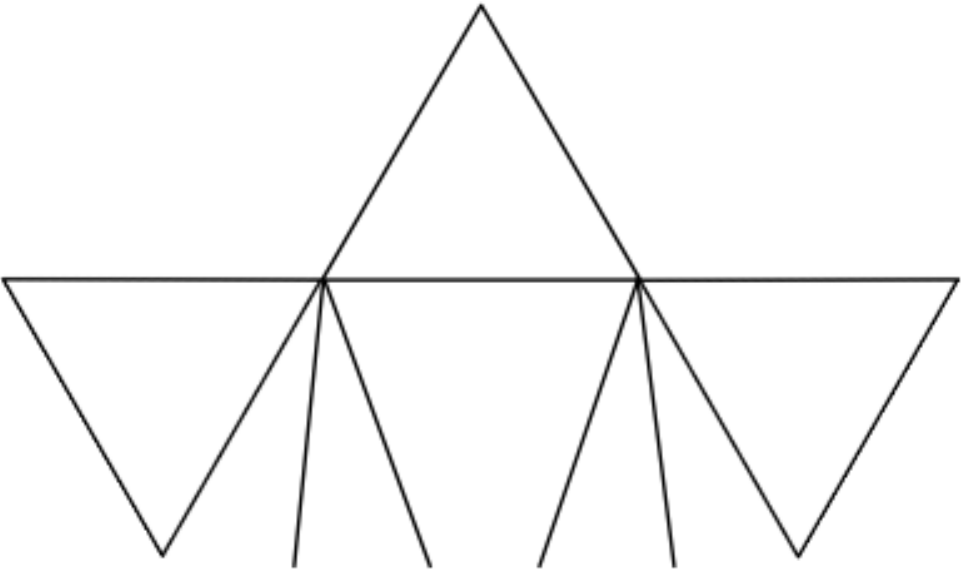}
}
\subfigure[]{
\includegraphics[width=0.4\columnwidth]{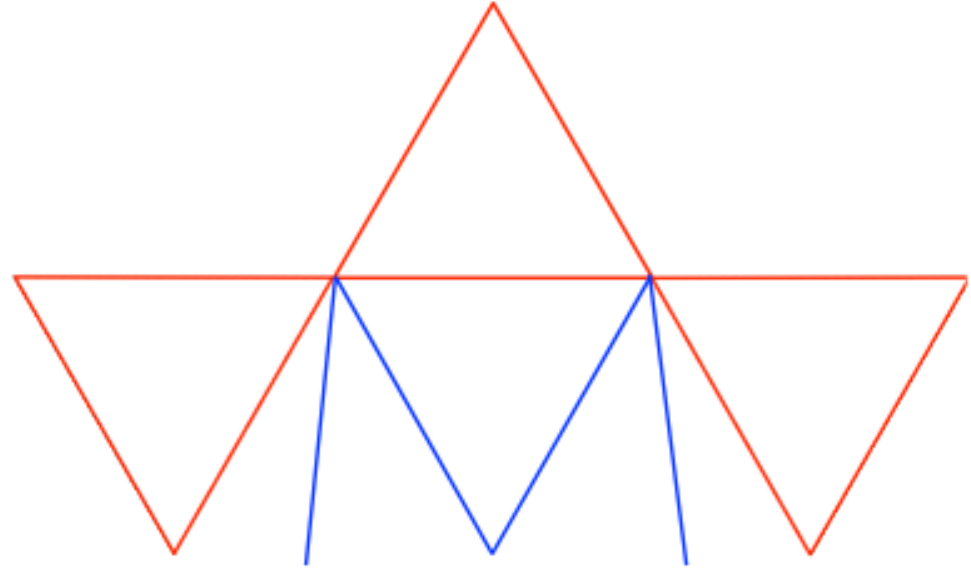}
}
\caption{(Color online) (a) A graph with no loop longer than $3$. (b) One way to decompose a graph containing loops longer than $3$ to triangles (red) and single edges (blue).  
	\label{fig:graph}
	}
\end{figure}

Given a general
network $G$ with $n$ nodes and $m$ edges, we can decompose the graph 
into set of triangles $\mathbb{T}$ and set of single edges $\mathbb{E}$ 
that do not share common edges, which we term the
$\{\mathbb{T},\mathbb{E}\}$ decomposition.
It is natural to define a factor graph composed of
$\mathbb{T}$ and $\mathbb{E}$ where triangles and single edges are
treated as two different kinds of factors, that is a factor graph
having both two-body interactions and three-body interactions, as in
the classic $2+p$ Satisfiability problem \cite{monasson1999determining}. 
An example of the decomposition is given in Fig.~\ref{fig:graph} (a).
In the figure the graph contains $11$ nodes, $3$ triangles and $4$ single edges, 
hence in the factor
graph there are $11$ variables and $7$ factors.
Then we apply Belief Propagation (BP) algorithm on this factor graph to compute
the marginal probability of each node being in the percolation cluster.
Then the percolation transition can be estimated using the marginals
which can be simplified to an eigenvector problem of a matrix, that we
call \textit{Triangle-Non-Backtracking} matrix. 

When the network contains no loops longer than $3$, as shown in the Fig.~\ref{fig:graph}, the $\{\mathbb{T},\mathbb{E}\}$
decomposition of the graph is unique, and marginals given by BP is exact
which results to an exact estimate of percolation transition.
However if network do contains loops longer than $3$, the $\{\mathbb{T},\mathbb{E}\}$ decomposition
is not unique. 
An example of choosing a 
$\{\mathbb{T},\mathbb{E}\}$ decomposition is illustrated in 
Fig.~\ref{fig:graph} where 
the graph has two triangles sharing an edge, thus there are two ways to
decompose the graph. In the figure only one way to do the 
decomposition is shown: putting the (red) upper triangle as a triangle
and putting blue edges as single edges.
%In this case we need to choose a $\{\mathbb{T},\mathbb{E}\}$ decomposition, and construct the factor graph. 
In this paper we do not address the problem of 
selecting the optimal $\{\mathbb{T},\mathbb{E}\}$ decomposition. Instead we note that for a 
randomly selected $\{\mathbb{T},\mathbb{E}\}$ decomposition, our approach gives a provable better estimate of
the percolation transition than the existing spectral bounds.

%\begin{figure}[h]
%  \centering
%%\subfigure[]{
%%{\includegraphics{fig2}}
%%}
%%\subfigure[]{
%%{\includegraphics{fig2}}
%%}
%\subfigure[]{
%\includegraphics[width=0.42\columnwidth]{fig3a.pdf}
%}
%\subfigure[]{
%\includegraphics[width=0.42\columnwidth]{fig3b.pdf}
%}
%\caption{(\textit{Left:}) A graph with no loop longer than $3$. (\textit{Right:}) One way to decompose a graph containing loops longer than $3$ to triangles (red) and single edges (blue).  
%	\label{fig:graph}
%	}
%\end{figure}
%

\section{Message Passing considering triangles}
In this paper we will address the bond percolation problem, the site percolation problem can be addressed using the analogous technique.
%Here we will use $n^-$ to indicate number of single edges and $n^\Delta$ to indicate number of triangles.
%	If three end-points of one triangle $a$ are $i, j$ and $k$,
%	then we will use $(i,j)\in a$ to indicate that edge $(i,j)$ belongs to triangle $a$.
For the bond percolation, each edge is open with probability $p$ and is
close with probability $1-p$. We are interested in size of cluster that 
each node belongs to $\{s_i\}:i\in [1,n]$ (where $s_i$ denotes the size of cluster that node $i$ belongs to, $n$ denotes the number of nodes). Since each realization of open-close configuration of
edges is a random variable, we can not predict $\{s_i\}$
for each realization. Instead we are interested in the probability of
node $i$ being in cluster of size $s$, $\psi_s^i$, the probability defined 
on all realizations of the open-close states of edges.
Particularly we are interested in the size of the giant
cluster $\hat s_1(p)$, which is an increasing function of $p$. 
For an infinite network, the 
percolation transition $p_c$ is defined at the critical point
where $\hat s_1(p)$ changes from
$0$ to a finite value, indicating that the giant percolation cluster
fills a non-vanishing fraction of nodes in the network. For a real-world network
which has a finite size, we are also interested in the size of 
the second largest percolation cluster $\hat s_2(p)$, as we can define $p^*$
at the point where the $\hat s_2(p)$ begins decreasing as $p$ increases.

We start with consistent equations for the size of \textit{finite} clusters that a node $i$
belongs to. 
When the graph has no loops longer than $3$, we can effectively treat the
factor graph as a tree rooted at $i$ and has two kinds of descendants: through
singles edges and through triangles. Clearly different descendants do not
share common nodes, thus size of the cluster that node $i$ belongs to,
$s_i$, is sum of cluster sizes of reachable (i.e. connected by a open edge) descendants. Then the 
probability that node $i$ belongs to a cluster with size $s$ can be written as
%\begin{equation}
%%	\label{eq:Bethe_Triangle}
$
\psi_{s_i}^i=\sum_{\{s_{i'}:i'\in \partial^- i\}}
\sum_{\{s_a:a\in \partial^\Delta i\}}\mydelta{s_i-1,\sum_{i'}s_{i'}+\sum_{a} s_a}
	\textrm{Prob}\bra{ \{s_{i'}\},\{s_a\}         },
$
%\end{equation}
where $\mydelta{a,b}$ is the Kronecker delta function, $\partial^- i$ denotes
the set of neighbors of node $i$ through a single edge, $\partial ^\Delta i$
denotes the set of triangles connected to node $i$, and 
$\textrm{Prob}\bra{ \{s_{i'}\},\{s_a\}}$ denotes the joint probability of
sizes reachable through triangles and single edges respectively. As the factor
graph is a tree, the joint probability of sizes can be written in a factorized form, which results to

%Let us first consider a simpler case where the graph is a 
%tree.
%On a tree, node $i$ belongs to cluster with size $s_i$ only when the sum of
%cluster size of reachable neighbors equals $s_i-1$, thus
%probability that node $i$ belonging to cluster with
%size $s_i$,
%$\psi_s^i$, satisfies that
%\begin{align}
%	\label{eq:Bethe_marg}
%	\psi_s^i&=\sum_{\{s_j:j\in \partial i\}}\mydelta{s_i-1,\sum_{j\in\partial i}s_j}
%	{\textrm{Prob}{s_j}}\nonumber\\
%	\psi_s^i&=\sum_{\{s_j:j\in \partial i\}}\mydelta{s_i-1,\sum_{j\in\partial i}s_j}
%{\prod_{j\in\partial i } \psi_{j\to i}^{s_j}},
%\end{align}
%where $\mydelta{a,b}$ is the Kronecker delta function, $\partial i$ denotes
%the set of neighbors of node $i$, and $\psi_{j\to i}^{s_j}$ is the
%(cavity) probability that node $j$ belongs to a cluster with size $s_j$
%reachable to node $i$ through edge $(i,j)$.
%Above relation is correct when neighbors of a node do not have other 
%common neighbors. When neighbors of a node $i$ do have other common neighbors
%for examples when the network contains triangles, say $(i,j,k)$, above relation is probably
%a bad approximation as it overcounts the size of clusters that $j$ and $k$
%both belong to. Clearly in this case a simple improvement to Eq.\eqref{eq:Bethe_marg} is to assume that 
%the cluster size of $i$ are contributed by clusters of neighbors of $i$
%reachable from single edges and triangles:
\begin{equation}
	\label{eq:Bethe_Triangle}
%$	
\psi_s^i=\sum_{\{s_{i'}:{i'}\in \partial^- i\}}
	\sum_{\{s_a:a\in \partial^\Delta i\}}\mydelta{s_i-1,\sum_{i'}s_{i'}+\sum_{a} s_a}
	{\prod_{i'\in\partial^- i } \psi^{{i'}\to i}_{s_{i'}}}
{\prod_{a\in\partial^\Delta i } \psi^{a\to i}_{s_a}}.
%$
\end{equation}
Here $\psi^{i'\to i}_{s_{i'}}$ denotes the 
probability that size of cluster that node $i'$ is reachable from $i$ through 
a single edge, and $\psi^{a\to i}_{s_a}$ is the probability that 
triangle $a$ belongs to a cluster with size $s_a$ that is reachable from node
$i$. 

If a single edge is close (with probability $1-p$) the size of cluster that
reachable through $i$ is clearly $0$, otherwise it must be a finite value.
Thus cavity probabilities $\psi_{i\to l}^{s_i}$ can be computed as
\begin{align}
%\begin{equation}
%	\label{eq:iter}
	\psi^{i\to l}_{s_i}=(1-p)\mydelta{s_i,0}+p\sum_{\{s_{a}:a\in\partial ^\Delta i\}}\sum_{ \{ s_{i'}:i'\in\partial^- i\backslash l\} }
	\mydelta{s_i-1,\sum_{i'}{s_{i'}}+\sum_{a}{s_a}}
	\prod_{i'}
	\psi_{s_{i'}}^{i'\to i} \prod_{a}\psi_{s_a}^{a\to i}. \nonumber
%\end{equation}
\end{align}
Messages sent from triangle $a=(i,j,k)$ to a node $i$, 
$\psi^{a\to i}_{s_a}$,
are more complex, as the probability depends on 
whether edges $(i,j), (j,k)$ and $(i,k)$ are open or close: If both $(i,j)$
and $(i,k)$ are close, with probability $(1-p)^2$, the number
of nodes reachable from node $i$ through triangle $a$, $s_a$ is $0$;
If both $(i,j)$ and $(i,k)$ are open, then node $j$ and $k$ are both reachable from node $i$,  
we can see that $s_a-1$ should equal to the number of nodes
in clusters that both $j$ and $k$ belong to; If only one of two
neighbors, say $j$,
is reachable from $i$, 
then $s_a-1$ should be the number of nodes in the cluster that $j$ belongs to.
Therefore we have the following equation for the probability
of number of nodes reachable from $i$ through triangle $a=(i,j,k)$,
\begin{align}
	\psi_{s_a}^{a\to i} &= 
	(1-p)^2 \mydelta{s_a,0}\nonumber\\
	&+p(1-p)^2\sum_{\{s_{j'}\}:j'\in\partial ^-j}\sum_{\{s_b\}:b\in\partial ^\Delta j \backslash a}\mydelta{s_a-1,\sum_{j'}s_{j'}+\sum_bs_b}\prod_{j'}\psi_{s_{j'}}^{j'\to j}
	\prod_{b}\psi_{s_{b}}^{b\to j} \nonumber\\
	&+p(1-p)^2\sum_{\{s_{k'}\}:k'\in\partial ^-k}\sum_{\{ s_f: \}f\in\partial ^\Delta k\backslash a}\mydelta{s_a-1,\sum_{k'}s_{k'}+\sum_fs_{f})}\prod_{k'}\psi_{s_{k'}}^{k'\to k}
	\prod_{f}\psi_{s_{f}}^{f\to k}
	\nonumber\\
	&+(3p^2-2p^3)
\sum_{\{s_{j'}\}:j'\in\partial ^-j}\sum_{\{s_b\}:b\in\partial j^\Delta \backslash a}
\sum_{\{s_{k'}\}:k'\in\partial ^-k}\sum_{\{ s_f \}f\in\partial k^\Delta \backslash a}
	\prod_{j'}\psi_{s_{j'}}^{j'\to j}
	\prod_{b}\psi_{s_b}^{b\to j}
\nonumber\\
	&\cdot\prod_{k'}\psi_{s_{k'}}^{k'\to k}
	\prod_{f}\psi_{s_f}^{f\to k}
	\mydelta{ s_a-1,\sum_{j'}s_{j'}+\sum_bs_b+\sum_{k'}s_{k'}+\sum_fs_f}\nonumber
\end{align}
We need to notice that the cluster size in the 
last equation must be \textit{finite}, as only if $\{s_j\}$ and 
$\{s_a\}$ are finite values, will the equality 
$s_i-1=\sum_{\{s_j:j\in \partial^- i\}}+\sum_{\{s_a:a\in \partial^\Delta i\}}$ make sense. Otherwise if cluster size of
at least one of $i$'s neighbors is $\infty$,
then $s_i$ must be $\infty$ regardless of the value of 
other cluster sizes.

\begin{figure}[h]
  \centering
\subfigure[]{
\includegraphics[width=0.4\columnwidth]{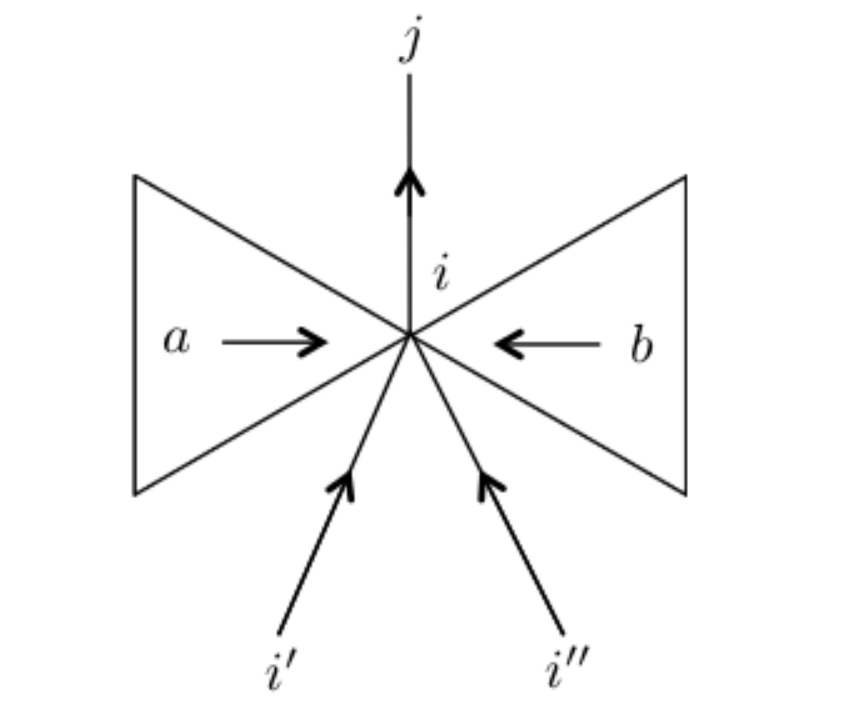}
}
\subfigure[]{
\includegraphics[width=0.48\columnwidth]{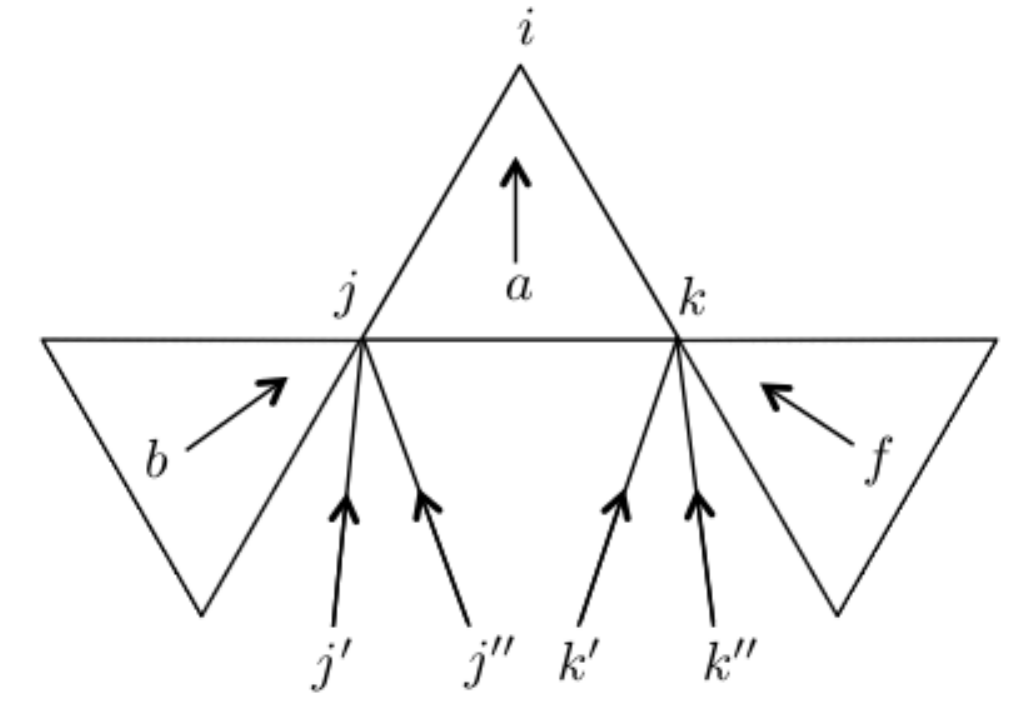}
}
	\caption{
	\label{fig:mp}
	(a): Illustration of the computation of BP messages along a
	single edge $i\to j$, using messages from triangles and single
	edges to $i$, except $j\to i$.
	(b): Illustration of computation of messages from 
	a triangle $a$ to node $i$, as a function of messages from 
	other triangles and from single edges.}
\end{figure}

We can see that equations above are difficult to solve, because the number
of states is large. So instead of trying to determine $\psi^i_{s_i}$, we 
introduce the total
probability that a node belongs to finite clusters
\begin{equation}
	\label{eq:bp:marg}
	\eta^i=\sum_{s_i=1}^\infty\psi^i_{s_i}=\prod_{i'\in\partial ^- i}\eta^{i'\to i}\prod_{a\in \partial^\Delta i}\eta^{a\to i},
\end{equation}
where $\eta^{i'\to i}$ and $\eta^{a\to i}$ are 
probabilities that $i'$ or $a$ belongs to finite clusters, and  will
be introduced later.
Eq.~\eqref{eq:bp:marg} actually has a simple meaning: 
\textit{A node belongs 
to a finite cluster only when all its reachable neighbors belong to 
finite clusters}. Then notice that the probability of a node $i$ belongs to 
an infinite (percolating) cluster is simply $1-\eta^i$. 
In the above equation the joint probability of its reachable neighbors
belonging to finite clusters has the factorized form because we 
assumed that the factor graph is locally-tree-like, and different neighbors
in different branches of the tree rooted at $i$ have no common children.
The probabilities of neighbors belonging to finite clusters can be
computed in a similar way for both along an edge $i\to j$ and from
a triangle $a=(i,j,k)$ to one of its end-point $i$, as illustrated in 
Fig.~\ref{fig:mp}:
\begin{align}
	\label{eq:bp:1}
	\eta^{i\to j}&=\sum_{s=0}^\infty\psi_s^{i\to j}
	=1-p+p\prod_{i'\in\partial i^-\backslash j}\eta^{i'\to i}\prod_{a\in\partial ^\Delta i}\eta^{a\to i}\\
\label{eq:bp:2}
	\eta^{a\to i}&=\sum_{s=0}^\infty\psi_s^{a\to i}
	=(3p^2-2p^3)\prod_{j'\in\partial ^- j}\eta^{j'\to j}
	\prod_{b\in\partial ^\Delta j}\eta^{b\to j}
	\prod_{k'\in\partial ^- k}\eta^{k'\to k}
	\prod_{f\in\partial ^\Delta k}\eta^{f\to k}\nonumber\\
	&+(1-p)^2+p(1-p)^2\bra{\prod_{j'\in\partial ^- j}\eta^{j'\to j}
	\prod_{b\in\partial ^\Delta j}\eta^{b\to j}+
	\prod_{k'\in\partial ^- k}\eta^{k'\to k}
	\prod_{f\in\partial ^\Delta k}\eta^{f\to k}}.
\end{align}
Above equations are belief propagation equations on the factor graph
associated with a $\{\mathbb{E},\mathbb{T}\} $ decomposition of the network. We can initialize BP messages randomly then
update messages using BP iterative equations in random sequential order.
After messages converge, the probability of each node being
in finite clusters 
can be computed using Eq.~\eqref{eq:bp:marg}.
We can also compute the expectation of fraction of nodes in the 
percolation cluster is given by 
%\begin{equation}
%	\label{eq:fraction}
$	\tilde s=\frac{1}{n}\sum_{i=1}^n(1-\eta^i).$
%\end{equation}
We note that our messages $\eta^{j\to i}$ and $\eta^{a\to i}$ correspond 
to a special case of generating functions used in \cite{karrer2014percolation}.
Other quantities, e.g. size of the non-percolating cluster that a node belongs
to, can be computed using the generating functions technique \cite{karrer2014percolation}, but will not be shown here.
It is straightforward to see that if for every node $i$ such that $\eta^i=1$ we have $\tilde s=0$. This means 
that every node belongs to finite clusters hence there is no
percolating cluster occupying finite fraction of nodes. When
$\tilde s>0$, i.e., for some $i$, $\eta^i$ deviates from $1$, then 
obviously system has a percolation cluster. 
Thus where $\eta^i$ deviate from $1$ tells us the position of 
the estimated percolation transition.

Notice that BP equations on the factor graph are exact only when the graph contains no
loops that are longer than $3$. 
On general graphs which do have loops longer than $3$, 
although the fact that finite-reachable nodes from 
neighbors leads to a finite-size cluster still holds, we can not 
express the joint probability that neighbors belonging to finite clusters
as a factorized form. Instead, for $i'\in\partial ^-i$ and
$a\in\partial ^\Delta i$ we need to use
\begin{equation}
	\eta^i=\textrm{Prob}\bra{\{s_{i'}\},\{s_a\} \textrm{ are finite}},
\end{equation}
and probabilities in Eq.~\eqref{eq:bp:1}~\eqref{eq:bp:2} need
to be reformulated as
\begin{align}
	\eta^{i\to j}&=
	1-p+p\cdot\textrm{Prob}\bra{\{s_{i'}\},\{s_a\} \textrm{ are finite}}\nonumber\\
	\eta^{a\to i}
	&=(1-p)^2+p(1-p)^2
\cdot\brb{\textrm{Prob}\bra{\{s_{j'}\},\{s_b\} \textrm{ are finite}}
+\textrm{Prob}\bra{\{s_{k'}\},\{s_f\} \textrm{ are finite}}}
	\nonumber\\
	&+(3p^2-2p^3)
\cdot\textrm{Prob}\bra{\{s_{j'}\},\{s_b\},\{s_{k'}\},\{s_f\} \textrm{ are finite}}.
\end{align}
Since above probabilities are positively correlated when
there are loops longer than $3$,
generally we have
%\begin{align}
$\textrm{Prob}\bra{\{s_{i'}\},\{s_a\} \textrm{ are finite}} \geq \prod_{i'}\eta^{i'\to i}\prod_{a}\eta^{a\to i}.$
%\end{align}
As a consequence, when BP equations converge, the obtained $\eta^i$ is only a lower-bound to the true probability
of node belonging to finite clusters, and deviates from $1$ earlier
than the true percolation transition as $p$ increases from a small value.
Thus percolation transition estimated using BP is only a lower-bound 
to the true percolation transition on an infinite size network.

\section{The Triangle-Non-Backtracking matrix}
We observe that $\eta^i=\eta^{i\to j}=\eta^{a\to i}=1$ is always a fixed-point of
BP Eqs.\eqref{eq:bp:1}\eqref{eq:bp:2}. This fixed-point means that all
nodes belong to non-percolation clusters so the size of largest cluster
is finite and the fraction of nodes that the percolation cluster
occupies is $0$. We call this fixed-point the \textit{factorized fixed-point}.
It is easy to check that at $p=0$ the factorized fixed-point
is the only fixed-point; with $p$ small the factorized fixed-point 
is locally stable to perturbations; 
when $p$ is large, the factorized fixed-point could be unstable during iteration
to an infinitesimal perturbations, and BP converges to another fixed-point
with a finite fraction of nodes occupied by the giant cluster. Therefore
the critical point $p^*$ where the factorized fixed-point becomes
unstable is our estimate of the percolation transition, and it can be studied by analyzing
the stability of the factorized fixed-point.

We expand BP messages at the factorized fixed-point:
\begin{align}
	\eta^{i\to j}&=1-\epsilon^{i\to j}&
	\eta^{a\to i}&=1-\epsilon^{a\to i},
\end{align}
then the iterative equations for the deviation of messages to the factorized fixed-point, 
along a directed edge $i\to j$ and from triangle $a=(i,j,k)$ to
one of its end-point $i$, can be written as
\begin{align}
	\label{eq:linear}
	\epsilon^{i\to j}&:=
p\sum_{i'\in\partial i^-\backslash j}
\epsilon^{i'\to i}
+p\sum_{a\in\partial ^\Delta i}\epsilon^{a\to i}\nonumber\\
\epsilon^{a\to i}&:=
q\sum_{j'\in\partial j^-}
\epsilon^{j'\to j}
+q\sum_{b\in\partial ^\Delta j\backslash a}\epsilon^{b\to j}
+q\sum_{k'\in\partial k^-} \epsilon^{k'\to k}
+q\sum_{f\in\partial ^\Delta k\backslash a}\epsilon^{f\to k},
\end{align}
with $q=p+p^2-p^3$.
Above equations can be rewritten as a matrix form
%\begin{equation}
	$\lambda\epsilon=C\epsilon,$
%\end{equation}
where $\lambda$ is an eigenvalue and $C$ is the matrix that we refer as \textit{Triangle-Non-Backtracking} (TNB) matrix,
whose elements $x\to i$ and $z\to w$ are messages along directed single edges or 
from a triangle to one of its end-point:
	\begin{equation}
		\label{eq:C}
		C_{x\to i,z\to w}=p\ind_{x\not\in \mathbb{T}}\mydelta{x,w}(1-\mydelta{z,i})+q\ind_{x\in\mathbb{T}}\ind_{w\in x\backslash i}\ind_{z\not \in x}.
	\end{equation}
Here $\ind$ is the indicator function. 
We can see that analogous to the non-backtracking matrix, 
the TNB matrix represents the non-backtracking
walks in the network, but avoids also the backtrackings along (weighted) 
edges of triangles.

We are interested in the largest eigenvalue $\lambda_C$ and the corresponding leading eigenvector 
$\epsilon_1$. Eq.~\eqref{eq:linear} offers a method for its evaluation.
As every element of matrix $C$ is non-negative, 
Perron-Frobenius theorem applies and we can conclude that the $\lambda_C\geq 0$ and 
elements of $\epsilon_1$ are non-negative.
$\lambda_C$ is actually a stability parameter for BP factorized 
fixed-point: When $\lambda_C<1$, the factorized fixed-point is stable,
system has no percolating cluster; when $\lambda_C>1$, system
has a percolating cluster occupying finite fraction of nodes. 
So $\lambda_C=1$ gives a critical 
percolation probability $p^*_C$ (note that $\lambda_C$ is an increasing function of $p$).
Furthermore, we establish in the following lemma that on an arbitrary graph, 
$\lambda_C$ is bounded above by $p\lambda_B$ when 
$p\lambda_B\le 1$, where $\lambda_B$ is
the leading eigenvalue of the non-backtracking matrix $B$. The proof to the lemma can be found in the appendices.

%\begin{theorem}
\begin{lemma}
	\label{lemma:1}
	Let $\lambda_B$ be the largest eigenvalue of the non-backtracking matrix, $\lambda_C$ be the largest eigenvalue of the triangle-non-backtracking matrix associated with any $\{\mathbb{T},\mathbb{E}\}$ decomposition of the network.
	Then for any undirected graph, we have $p\lambda_B\geq \lambda_C$ when $p\lambda_B\leq 1$.
%\end{theorem}
\end{lemma}

The above lemma says that below the percolation transition given by 
$p^*\lambda_B=1$, $\lambda_C$ never excesses $1$. With the fact
that $\lambda_C$ is an increasing function of $p$, we conclude that
the critical probability that makes $\lambda_C=1$ is bounded below
by $p^*\lambda_B$. Also by making use of the theorem in \cite{karrer2014percolation,Hamilton14}
that the leading eigenvalue of the adjacency matrix, $\lambda_A$ is strictly
larger than $\lambda_B$, we prove the following theorem.
\begin{theorem}
	\label{theo:1}
	Let $p^*_C$ satisfy $\lambda_C=1$ for the triangle-non-backtracking matrix
	$C$ associated with an arbitrary $\{\mathbb{T},\mathbb{E}\}$ decomposition of
	the graph, let $p_B^*=\frac{1}{\lambda_B}$, $p_A^*=\frac{1}{\lambda_A}$,
then on an arbitrary undirected graph, 
$p_A^*< p_B^*\leq p_C^*$.
\end{theorem}
So, in general, the percolation transition making the leading eigenvalue
of the triangle-non-backtracking matrix equal $1$ is a tighter 
lower bound to the true percolation transition than
the lower-bound given by the inverse of the leading eigenvalue of the non-backtracking matrix and the adjacency matrix.

If we use $n^-$ to indicate the number of single edges and $n^\Delta$ indicate number of triangles, then the TNB matrix Eq.~\eqref{eq:C} has size $2n^-+3n^\Delta$, which 
becomes larger when network becomes denser. Fortunately we see
that the non-trivial part of spectrum of C can be obtained from a 
matrix $C'$ (as defined in Definition~\ref{def:1}) with size $4n\times 4n$ --- which does not increase with number
of edges. This statement is established
in Theorem~\ref{theo:2} and is proved in the appendices.
\begin{definition}
	\label{def:1}
	Define matrix $C'$ that associated with a $\{\mathbb E,\mathbb{T}\}$ 
	decomposition as
\begin{align}
C' = \begin{pmatrix}
p \A^- & -p \id & p \A^- & 0 \\
p (\D^- - \id) & 0 & p \D^- & 0 \\
q \A^\Delta & 0 & q \A^\Delta & -q \id \\
2q \D^\Delta+q \A^\Delta & 0 & 2q(\D^\Delta-\id)+q \A^\Delta & -q \id
\end{pmatrix} \, . 
\end{align}
\end{definition}
Here $\id$ denotes the $n$-dimensional identity matrix, $\A^-$ is
the adjacency matrix of nodes connected by single edges; $\D^-$ is
the diagonal matrix of single-edge-degrees;$\A^\Delta$ is 
the adjacency matrix of nodes connected by triangles; $\D^\Delta$
is the diagonal matrix of triangle-degrees. That is
\begin{align}
	\A^-_{i,j} &= \begin{cases} \mbox{$1$, if $(i,j)\in\mathbb{E}$} \\ \mbox{$0$, other wise}  \end{cases}& 
%		\D^-_{i,j} &= \begin{cases} \mbox{$\sum_k\A^-_{ik}$ if $i=j$} \\ \mbox{$0$ other wise}  \end{cases} \\
\A^\Delta_{i,j} &= \begin{cases} \mbox{$1$, if $\exists (i,j,l)\in\mathbb{T}$ }\\ \mbox{$0$, other wise}  \end{cases} \\
%		\D^\Delta_{i,j} &= \begin{cases} \mbox{$\frac{1}{2}\sum_k\A^\Delta_{ik}$ if $i=j$} \\ \mbox{$0$ other wise}  \end{cases} \\
		\D^-_{i,i} &=\sum_k\A^-_{ik}&
		\D^\Delta_{i,i} &=\frac{1}{2}\sum_k\A^\Delta_{ik}.
\end{align}

\begin{theorem}
	\label{theo:2}
	If $C$ and $C'$ are associated with the same $\{\mathbb T,\mathbb{E}\}$ 
	decomposition of an arbitrary graph, then non-zero eigenvalues of two matrices are identical.
\end{theorem}
Thus working with this $4n$ dimensional matrix can significantly reduce
the computational complexity of finding the leading eigenvalue of
matrix $C$.
Obviously when the network has no loops longer than $3$, that is the 
factor graph treating both single edges and triangles as factors is 
a tree, $p^*_C$ is exact. It is interesting to see that when
the factor graph is a finite tree, $\lambda_C=0$. This is because when
we can find the leaves in the factor graph, the elements in the matrix $C$ which corresponds to the single edges connected to the leaves
must be $0$. The $0$ messages will
be propagated further to all the network, resulting to $\lambda_C=0$. 
This do 
makes sense in the finite-size networks, as there could not be
infinite-size cluster or percolation transition defined in the system.
Also note that when the network is infinite, where it is not possible 
to define the concept of a leave, $\lambda_C>0$ due to proper boundary conditions.
At the thermodynamic limit the (asymptotic) exactness of our method does not require the factor graph to be a perfect tree. Instead, when the fraction of short loops in the factor graph is negligible, in other words when correlations decays fast enough~\cite{Krzakala2007} our method would be asymptotically exact. Similar arguments on the asymptotically exactness have been used in~\cite{karrer2014percolation}, as well as in the context of community detection in networks~\cite{Decelle2011a}.

When the infinite network do have loops longer than $3$, 
the resulting factor
graph is no longer a tree, then our estimate $p^*_C$ on any 
$\{\mathbb{T},\mathbb{E}\}$ decomposition of single edges and triangles is a lower bound 
for the true percolation transition, and is tighter than the inverse
of the non-backtracking matrix.

\section{Experimental Evaluation}
In this section, to evaluate our message passing algorithm, as well as
the estimate of percolation transition using the TNB matrix, we conduct 
direct simulations of percolation processes on 
both synthetic and real-world networks and compare the average
size of the giant cluster and the percolation transition with our methods. 

\subsection{Synthetic networks}
For synthetic 
networks we use the \textit{Clustered Random Graph} (CRG)
model proposed in \cite{newman2009random}. In the CRG model, 
the standard random graph models are generalized to incorporate triangles.
The model has a tunable clustering coefficient by specifying both number
of triangles and number of single edges then distributing these edges
and triangles into networks in a random and uncorrelated way.
In our simulations we
use parameter $\rho$ to specify number of edges belonging to triangles.

\begin{figure}[h]
  \centering
\subfigure[]{
\includegraphics[width=0.42\columnwidth]{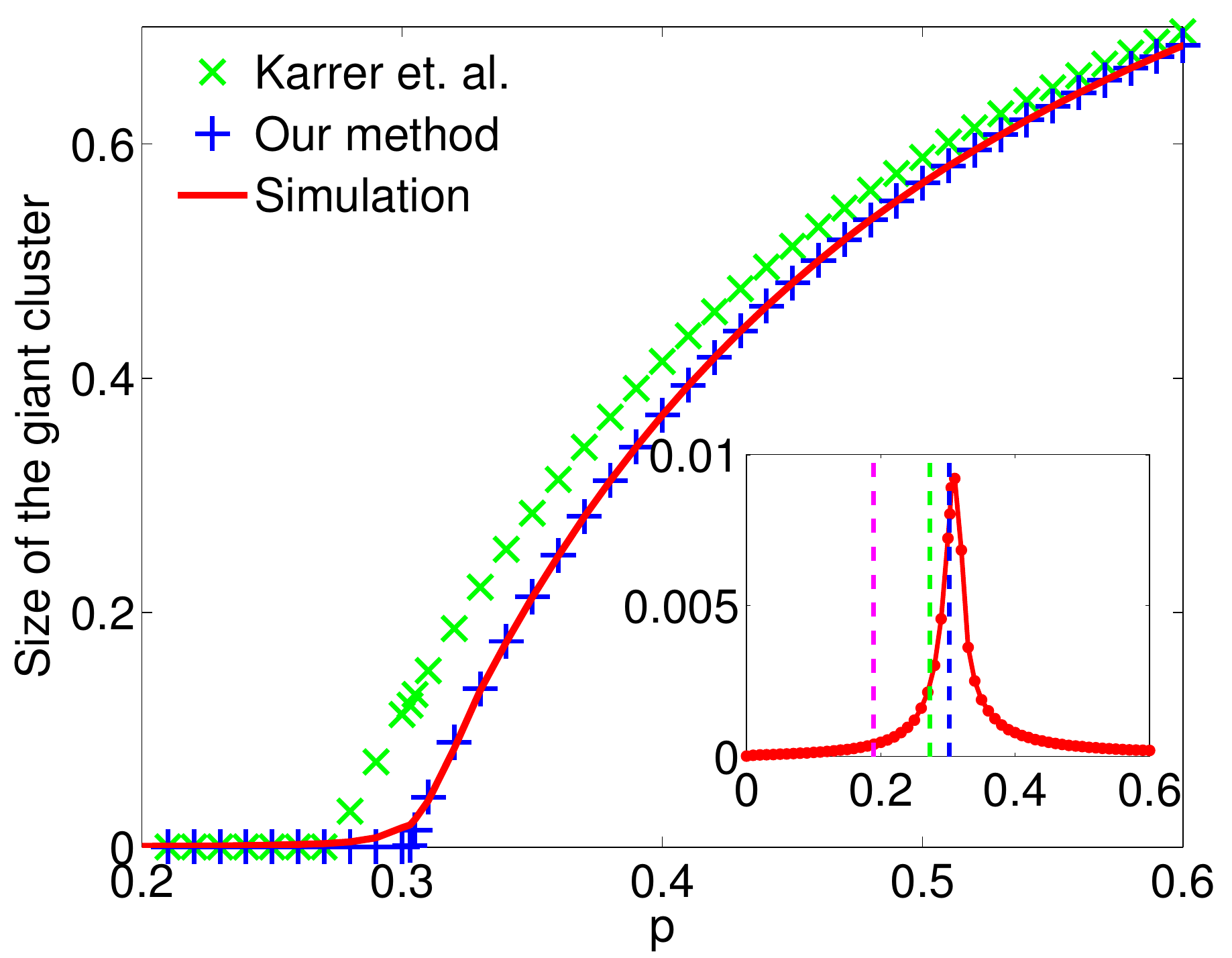}
}
\subfigure[]{
\includegraphics[width=0.42\columnwidth]{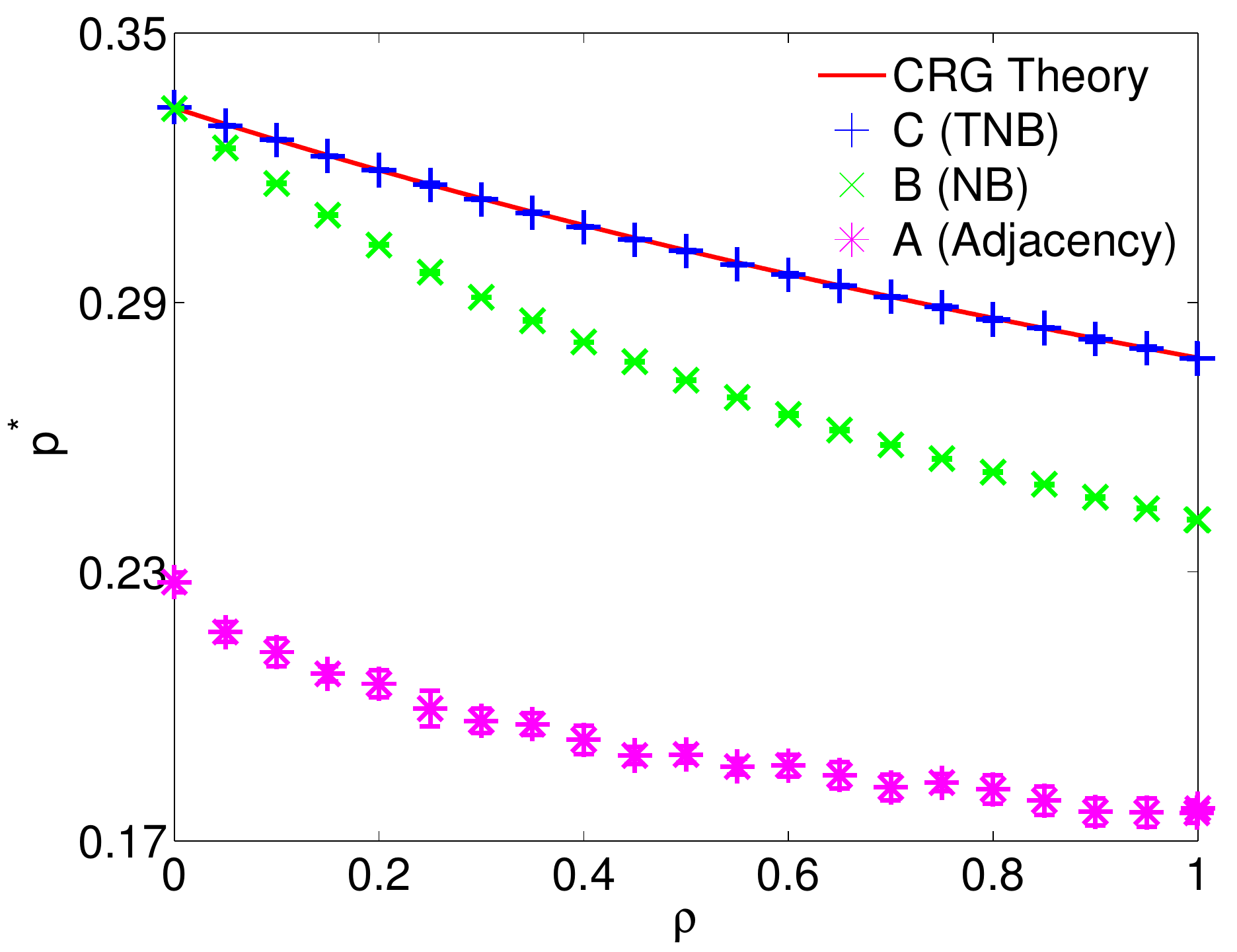}
}
\caption{
(Color online) 
	\label{fig:crg:1}
	(a) Comparison of size of the giant clusters of
simulations (averaged over $10$ realizations), BP without triangles 
\cite{karrer2014percolation} and our message passing algorithm on
a network generated by the clustered random graph model \cite{newman2009random}. 
The network has number of nodes $n=10^5$, average degree $c=3$, and fraction of edges
belonging to triangles $\rho=0.5$.
The inset shows the size of the second largest cluster in simulations which
has a peak at the percolation transition. Three vertical lines,
from left to right, in the 
inset are the estimates of percolation transition given using
the adjacency matrix, the non-backtracking matrix and the triangle-non-backtracking matrix respectively.
(b) From bottom to top, percolation transition given by the adjacency matrix $A$, 
the non-backtracking matrix $B$, the triangle-non-backtracking matrix $C$,
and exact theory for $n\to\infty$ CRG networks in \cite{newman2009random}.
Networks have a Poisson degree distribution with average degree $c=3$. 
Number of nodes $n=10^6$, $\rho$ denotes fraction of edges belonging to triangles.
Each point of simulation is averaged over $5$ 
random realizations of edges' open-close states.
}
\end{figure}

Since we need to test
the performance of our message passing algorithm and the
threshold given by the triangle-non-backtracking matrix, we conduct 
simulations on generated finite networks which do have loops longer than
$3$. So we first choose a (random) decomposition of the generated network into
single edges and triangles randomly, then run our BP algorithm and
triangle-non-backtracking matrix on the resulting graph.
We compare the size of the giant cluster in simulation with 
that computed using our method, and the state-of-art method for sparse
networks proposed in \cite{karrer2014percolation}
which does not incorporate effects of triangles.
The comparison is shown in Fig.~\ref{fig:crg:1}
for CRG networks with a Poisson degree distribution. 
From the figure we can see that the size of the giant cluster given by 
our method is in agreement with the simulation, while the message passing without
considering triangles as proposed in \cite{karrer2014percolation} overestimates
the average size of the giant cluster, hence underestimates the
percolation transition.
In the inset of the figure we plot the percolation transition estimated
using the adjacency matrix, the
non-backtracking matrix, and the triangle-non-backtracking matrix, as
well as the second largest component size, which displays a peak at
$p^*$ which can be seen as an estimate of percolation transitions in
a finite network. As the figure shows, the threshold given by the
triangle-non-backtracking matrix is much closer to $p^*$ than
the adjacency matrix and the non-backtracking matrix. All three estimates
are smaller than $\rho^*$, which is consistent with the argument that
they are all lower bounds for the true percolation transition on
an infinite network.

In the CRG model, the exact properties of bond percolation with $n\to \infty$
has already been
calculated \cite{newman2009random}. So we only need to test whether the 
thresholds given by the triangle-non-backtracking matrix on large network agree with
the $n\to\infty$ theory in \cite{newman2009random}. 
In Fig.~\ref{fig:crg:1} we compare thresholds given by 
the adjacency matrix $A$, the non-backtracking matrix $B$ and 
the triangle-non-backtracking matrix $C$ on large CRG networks with 
$n=10^6$ nodes, average degree $c=3$ and varying fraction
of edges belonging to triangles $\rho$, with the theoretical results for $n\to\infty$ in
\cite{newman2009random}. As the figure shows, the threshold given
by $C$ agrees very well with the $n\to\infty$ theory while thresholds
given by $B$ and $A$ are much worse.

\subsection{Real world networks}
In Table~\ref{tab:compare} we compare percolation transitions estimated 
using different matrices with simulations, on several networks 
with size ranging from $34$ nodes to $10^5$ nodes. The first $10$ networks 
in the table are real-world networks with references given in the table. The last two networks are synthetic networks with community structures, they are generated by the 
Stochastic Block Model (SBM) and its variant Triangular Stochastic Block Model (TSBM [unpublished]) which generates random graphs with both local (triangles) and global clusters (communities).
Since there is no real phase transition in finite-size networks, for simulation results $p_\textrm{simu}^*$ are taken at the point where the second largest cluster is greatest. The number in the parentheses indicate the error on the last digit.
The columns $p^*_A$, $p^*_B$, and $p^*_C$ refer to the percolation transition estimated using the leading eigenvalue of the adjacency matrix, non-backtracking matrix and the triangle-non-backtracking matrix respectively. 
We can see from the table that the errors of $p_C^*$ (number in parentheses) over different $\{\mathbb{T},\mathbb{E}\}$ decompositions are quite small. 
In agree with our theoretical justifications, our results shows that thresholds estimated using 
the eigenvalues of triangle-non-backtracking matrix are always lower than 
the simulation results, and are higher than those obtained using
the non-backtracking matrices and adjacency matrices.
For some networks such as the Karate club and TSBM networks, the improvement of $p_C^*$ over $p_B^*$ is quite large, while on networks which contains almost no triangles (such as the SBM network), $p_C^*$ is very close to $p_B^*$. 

For the SBM network, we can see a little difference between $\rho^*_B$ and $\rho^*_C$, which is clearly due to existing of few triangles ($6$ in this instance of SBM network). There are also a significant difference between $\rho^*_\textrm{simu}$ and $\rho^*_B$. We claim it should come from the finite size effect of using the point where the second-largest cluster is greatest as the estimate to the phase transition. With system size increases, first we have checked that both $\rho^*_B$ and $\rho^*_C$ converge to $0.333$, as the average excess degree (which equals to the average degree in with Poisson degree distribution) is $3.0$; then we confirm that $p^*_\textrm{simu}$ decreases with $n$ increasing. For example when network size increases to $5\times 10^5$,  $p^*_\textrm{simu}$ decreases to $0.338(1)$. Thus we expect $\rho^*_B$, $\rho^*_C$, and $\rho^*_\textrm{simu}$ coincide with network size $n\to\infty$ for SBM network, and expect $\rho^*_C$ and $\rho^*_\textrm{simu}$ coincide for the same reason.

\begin{table}[t]
  \centering
\begin{tabular}{c| c c c c c c c}
	\hline
	Network & $n$ & $m$ & CC & $p^*_A$ & $p^*_B$ &$p^*_C$ &$p^*_{\textrm{Simu}}$ \\
	\hline
	Karate club \cite{Zachary1977}&34&78 & 0.571 & 0.149 & 0.189 &0.197(1) & 0.229(5)\\
    Les Miserables \cite{Knuth93}&77&254&0.573 & 0.0833 & 0.0930 & 0.09432(3)&0.146(6) \\
	Political books \cite{politicalbooks}&105&441&0.488 & 0.0838 & 0.0941 & 0.0954(1)& 0.169(2) \\
	Political Blogs \cite{Adamic2005}&1490&16716&0.263 & 0.0135 & 0.0137 &0.0138(0) & 0.0173(6)\\
	Coauthorships \cite{Newman2006} &1589 &2742&0.638& 0.0526 &0.0555  &0.0558(0) &0.45(1) \\
	Protein \cite{jeong2001lethality} &2115&2203&0.0594& 0.133 & 0.198 & 0.206(1)& 0.31(1)\\
	Power grid \cite{Watts1998} &4941&6594&0.08& 0.134 & 0.161 &0.166(1) & 0.659(7)\\
	Collaboration \cite{newman2001structure}&8361&15751&0.42 & 0.0435 & 0.0454 & 0.0456(0)& 0.121(3)\\
	Email \cite{Leskovec2009}&36692&183831&0.497 & 0.00844 & 0.00866 & 0.00866(0)& 0.0104\\
	Gnutella \cite{matei2002mapping}&62586&147892&0.00546 & 0.0759 & 0.0871 &0.0872(0) & 0.0967\\
	SBM \cite{holland1983stochastic}&$100 000$&149998& $3.36\times 10^{-5}$&0.229&0.3327&0.3328(0)&0.341(1)\\
	TSBM [unpublished]&$100 000$&$99992$ &$0.272$&0.200&0.331&0.429(0)&0.444(2)\\
	\hline
  \end{tabular}
	\caption{
	\label{tab:compare}
	Percolation transitions estimated using eigenvalues of the adjacency
	matrices $p^*_A$, non-backtracking matrices $p^*_B$, 
	triangle-non-backtracking matrices $p^*_C$, and in 
	direct simulations $p^*_\textrm{simu}$ which is estimated by finding the point where the second-largest cluster is greatest, by taking average over at least $1000$ 
	realizations of percolation process. Each $p^*_C$ value is averaged over$10$ different $\{\mathbb{T},\mathbb{E}\}$ decompositions, numbers in parentheses indicate the error on the last digit. In each row, $n$ denotes the number
	of nodes, $m$ denotes the number of edges and CC indicates the clustering
	coefficient.
	}
  \end{table}

%\begin{figure*}
%\begin{center}
%\subfigure[]{\includegraphics[width=0.45\textwidth]{fig1a}}
%\qquad
%\subfigure[]{\includegraphics[width=0.45\textwidth]{fig1b}}
%\end{center}
%\caption{(Color online) The overlap $Q$, Eq.~\eqref{eq:ovl}, and the marginal overlap $Q_\mu$, Eq.~\eqref{eq:movl}, for belief propagation on networks generated by the stochastic block model with $q=2$ groups, $n=10^5$ nodes, average degree $c=3$, and group sizes as given in Eqs.~\eqref{eq:delta} and~\eqref{eq:zeta} a function of $\epsilon=\cin-\cout$ for various values of $\delta$.  Increasing $\delta$ (from bottom to top at the left of both panels) corresponds to greater differences between the group sizes and average degrees.  The dashed lines in the left panel are the expected values in the weak-structure (i.e.,~$\epsilon=0$) limit, Eq.~\eqref{eq:ovlweak}.  Note how the sharp detectability transition disappears for $\delta > 0$; both overlaps are smooth functions of the block model parameters.}
%\label{fig:q2ovl}
%\end{figure*}

\section{Conclusions and discussions}
\label{sec:conclusions}
This work extends and generalizes the recently proposed 
spectral bounds for the bond percolation to
incorporate effects of triangles in the network. We propose 
a message passing algorithm for computing the size of percolating
cluster considering triangles in the network, and demonstrate that
the linearized version of this message passing algorithm, which uses
the triangle-non-backtracking matrix, gives a tighter lower bounds 
than using other matrices, to the percolation transition on an arbitrary network.
On synthetic networks generated by the clustered random graph model we 
have shown that our estimate of the percolation transition on large
networks agrees with
the exact results for infinite networks. On real-world networks, we 
have shown that our estimate is always closer to the result of
direct simulations than the estimates given by other spectral methods.

Incorporating effects of triangles is the first step towards understanding
the effects of clustering to the percolation transitions. 
In principle our method can be generalized to include higher-order
structures like quadrangles and pentagons.
Without loss of generality, we can think about a structure like quadrangle.
In this case we will be having a factor graph of $3$ factors --- single edges,
triangles and quadrangles. By applying BP equations to this factor graph,
then taking the linearization at the factorized fixed point, we will be 
arriving at a matrix that is a generalized version of our triangle-non-backtracking
matrix, but considering quadrangles. 
Although the generalized matrix will be more and more complex when we 
consider higher and higher order structures, the gain is obvious: it gives a sequence
of lower-bounds to the true percolation transition.
We will put it in future study to test
whether considering higher-order structures improves significantly 
the estimate of the transition for real-world networks.

An interesting application of percolation is the network attack, or network dismantling problem. The problem asks to find the smallest set of nodes such that after their removal the size of the largest connected component is sub-extensive. The most efficient and effective methods for this problem are based on non-backtracking matrix ~\cite{Morone2015,Zdeborova2016}. These methods essentially ask a fundamental question on how size of the LCC changes if one removes each node individually. It turns out that this question is difficult to answer by running direct simulations, but quite easy to study using spectral methods, which convert the problem to the shift of leading eigenvalue under perturbations of node removal. It would be interesting to study whether spectral dismantling methods based on the generalized non-backtracking matrix significantly outperforms the existing methods that based on the non-backtracking matrix.

Finally it worths noting that the $4n\times 4n$ form of the triangle non-backtracking matrix has a similar form to the non-backtracking operator defined for dynamic networks proposed in~\cite{PhysRevX.6.031005}. It would be interesting to study the detailed relations between these two operators.

A c++ implementaiton of our algorithm can be found at {http://lib.itp.ac.cn/html/panzhang/perc.zip}
\begin{acknowledgments}
We acknowledge helpful suggestions from the anonymous referee.
Part of the computations have been carried out at the High Performance Computational Cluster of ITP, CAS.
\end{acknowledgments}
\appendix
	\section{Proof of Lemma 1}
	If the network contains no triangle, obviously the equality holds. 
	For networks with triangles, observe that in any $\{\mathbb{T},\mathbb{E}\}$
	decomposition, all single edges are single edges in the original graph, thus
	the matrix $C$ corresponding to the decomposition has a smaller size than
	the matrix $B$.
	Note that the leading eigenvector $\{u_{i\to j}\}$ 
	,with eigenvalue $\lambda_B$, of the non-backtracking matrix~\cite{Krzakala2013}  $$B_{i\to l,j\to k}=\delta(k,i)(1-\delta(j,l))$$ 
	satisfies
	\begin{align}
		u_{i\to l}&=\frac{1}{\lambda_B}\sum_{j\in\partial i\backslash l}u_{j\to i},\nonumber
	\end{align}
	we can rewrite the last equation in a $\{\mathbb{T},\mathbb{E}\}$ 
	decomposition as
	\begin{align}
		\lambda_B u_{i\to l}&=\sum_{j\in\partial^- i\backslash l}u_{j\to i}
		+\sum_{(i,j,k)\in\partial^\Delta i}(u_{k\to i}+u_{j\to i}),\nonumber\\
		&=\sum_{j\in\partial^- i\backslash l}u_{j\to i}
		+\sum_{(i,j,k)\in\partial^\Delta i}u_{(j,k)\to i}.
	\end{align}
	Here $u_{(j,k)\to i}$ represents the sum of messages from two
	other end-points in the triangle $(i,j,k)$, and it can be further evaluated as
	\begin{align}
	\lambda_B u_{(j,k)\to i}&=
\sum_{j'\in\partial ^-j}u_{j'\to j}+\sum_{(j,r,s)\in\partial^\Delta j\backslash (i,j,k)}u_{(r,s)\to j}	+\sum_{k'\in\partial ^-k}u_{k'\to k}+\sum_{(k,r,s)\in\partial^\Delta k\backslash (i,j,k)}u_{(r,s)\to k}
	+u_{j\to k}+u_{k\to j}\nonumber\\
	&=\sum_{j'\in\partial ^-j}u_{j'\to j}+\sum_{(j,r,s)\in\partial^\Delta j\backslash (i,j,k)}u_{(r,s)\to j}
	+\sum_{k'\in\partial ^-k}u_{k'\to k}+\sum_{(k,r,s)\in\partial^\Delta k\backslash (i,j,k)}u_{(r,s)\to k}\nonumber\\
	&+\frac{1}{\lambda_B}\bra{\sum_{j'\in\partial ^-j}u_{j'\to j}+\sum_{(j,r,s)\in\partial^\Delta j\backslash (i,j,k)}u_{(r,s)\to j}+u_{i\to j}+\sum_{k'\in\partial ^-k}u_{k'\to k}+\sum_{(k,r,s)\in\partial^\Delta k\backslash (i,j,k)}u_{(r,s)\to k}+u_{i\to k}}\nonumber\\
	&=(1+\frac{1}{\lambda_B})\left ( \sum_{j'\in\partial ^-j}u_{j'\to j}+\sum_{(j,r,s)\in\partial^\Delta j\backslash (i,j,k)}u_{(r,s)\to j}\right .\nonumber\\
	&\left . +\sum_{k'\in\partial ^-k}u_{k'\to k}+\sum_{(k,r,s)\in\partial^\Delta k\backslash (i,j,k)}u_{(r,s)\to k}\right )
	+\frac{1}{\lambda_B}(u_{i\to k}+u_{i\to j}).
	\end{align}
	
	Then we define a new matrix $B'$ with size $(n^- +n^{\Delta})\times (n^- +n^{\Delta})$, as
	\begin{align}
		B'_{x\to i,z\to w}&=p\ind_{x\not\in \mathbb{T}}\mydelta{x,w}(1-\mydelta{z,i})
		+p(1+\frac{1}{\lambda_B})\ind_{x\in\mathbb{T}}\ind_{w\in x\backslash i}\ind_{z\not\in x}
		+\frac{p}{\lambda_B}\ind_{x\in\mathbb{T}}\ind_{w\in x\backslash i}\mydelta{z,i}.
	\end{align}
	We can see that the leading eigenvalue of $pB$ and $B'$ are the same. 

	Since matrix $C$ is defined as (see main text):
	\begin{equation}
		\label{eq:C}
		C_{x\to i,z\to w}=p\ind_{x\not\in \mathbb{T}}\mydelta{x,w}(1-\mydelta{z,i})+q\ind_{x\in\mathbb{T}}\ind_{w\in x\backslash i}\ind_{z\not \in x}.
	\end{equation}
	It is easy to see that $B'$ and $C$ have the same size. 
	With $q=1+p^2-p^3$ we have
	\begin{align}
		&B'_{x\to i,z\to w}-C_{x\to i,z\to w}=
		(\frac{p}{\lambda_B}-p^2+p^3)\ind_{x\in\mathbb{T}}\ind_{w\in x\backslash i}\ind_{z\not\in x}
		+\frac{p}{\lambda_B}\ind_{x\in\mathbb{T}}\ind_{w\in x\backslash i}\mydelta{z,i}.
	\end{align}
As $p\geq 0$, the last term in the right hand side of last equation is always non-negative. 
We can also see that together with many other conditions, under condition
 $p\lambda_B\leq 1$,
we have 
$$\frac{p}{\lambda_B}=\frac{p^2}{p\lambda_B}\geq p^2\geq p^2-p^3,$$
thus every element of $B'-C$ is non-negative.
	By applying the Collatz-Wielandt theorem, the leading eigenvalue of $B'$ satisfies
	$$p\lambda_B\geq \min_{x\to y,v_{x\to y}\neq 0}\frac{(B'v)_{x\to y}}{v_{x\to y}}.$$
	where $v_{x\to y}$ can be any real vector with length $n^-+n^\Delta$. 
	If we choose $v$ to be the leading eigenvector of $C$, whose elements are non-negative,
	then under $p\lambda_B\leq 1$, we have
%	\begin{align}
%		\lambda_B&\geq\min_{x\to y,v_{x\to y}\neq 0}\bra{\frac{\brb{Cv}_{x\to y}}{v_{x\to y}}+\frac{\brb{(B'-C)v}_{x\to y}}{v_{x\to y}}}\nonumber\\
%		&\geq \lambda_C. \nonumber
%	\end{align}
	\begin{align}
		p\lambda_B\geq\min_{x\to y,v_{x\to y}\neq 0}\bra{\frac{\brb{Cv}_{x\to y}}{v_{x\to y}}+\frac{\brb{(B'-C)v}_{x\to y}}{v_{x\to y}}}
		\geq \lambda_C. 
%		\qquad\qquad\qquad\qquad
	\end{align}
The proof is complete.\hfill\ensuremath{\square}

\section{Proof of Theorem 4}
Eigenvector $\epsilon$ associated with a non-zero eigenvalue $\lambda$ of matrix $C$ satisfies
\begin{align}
	\label{eq:app:1}
	\lambda\epsilon^{i\to j}
&=p\sum_{i'\in\partial i^-\backslash j}
\epsilon^{i'\to i}
+p\sum_{a\in\partial ^\Delta i}\epsilon^{a\to i}\nonumber\\
\lambda\epsilon^{a\to i}
&=q\sum_{j'\in\partial j^-}
\epsilon^{j'\to j}
+q\sum_{b\in\partial ^\Delta j\backslash a}\epsilon^{b\to j}
+q\sum_{k'\in\partial k^-} \epsilon^{k'\to k}
+q\sum_{f\in\partial ^\Delta k\backslash a}\epsilon^{f\to k},
\end{align}
where triangle $a=(i,j,k)$.

Let us define sum of incoming messages and outgoing messages 
for node $i$ along edges connected to node $i$ as
\begin{align}
	\label{eq:app:2}
	\ein^{i-}=\sum_{i'\in\partial^-i}\epsilon^{i'\to i}& &
	\eout^{i-}=\sum_{i'\in\partial^-i}\epsilon^{i\to i'}.
\end{align}
Sum of incoming messages and outgoing messages for node $i$
along triangles connected to node $i$ can be defined in a similar way:
\begin{align}
	\label{eq:app:3}
	\ein^{i\Delta}=\sum_{a\in\partial^\Delta i}\epsilon^{a\to i}& &
\eout^{i\Delta}=\sum_{a\in\partial ^\Delta i}\sum_{j\in a\backslash i}\epsilon^{a\to j}
%	\eout^{i\Delta}=\sum_{i'\in\partial^\Delta i}\epsilon^{i\to i'}.
\end{align}
Inserting Eq.~\eqref{eq:app:1} into Eq.~\eqref{eq:app:2} we have
an expression for sum of incoming messages
along single edges to node $i$:
\begin{align}
	\label{eq:app:2}
	\lambda\ein^{i-}&=\lambda\sum_{i'\in\partial^-i}\epsilon^{i'\to i}\nonumber\\
	&=p\sum_{i'\in\partial^-i}\bra{\sum_{i''\in\partial i'^-\backslash i}
\epsilon^{i''\to i'} +\sum_{a\in\partial ^\Delta i'}\epsilon^{a\to i'}}\nonumber\\
	&=p\sum_{i'\in\partial^-i}\bra{\sum_{i''\in\partial i'^-}
\epsilon^{i''\to i'} +\sum_{a\in\partial ^\Delta i'}\epsilon^{a\to i'}
-\epsilon^{i\to i'} }\nonumber\\
%&=q\sum_{i'\in\partial^-i}\bra{\ein^{i'-}+\ein^{i'\Delta}
%-\epsilon^{i\to i'} }\nonumber\\
&=p\sum_{i'\in\partial^-i}\bra{\ein^{i'-}+\ein^{i'\Delta}}
-p\eout^{i-},
\end{align}
and also for sum of outgoing messages along single edges from node $i$:
\begin{align}
	\lambda\eout^{i-}&=\lambda\sum_{i'\in\partial^-i}\epsilon^{i\to i'}\nonumber\\
	&=p\sum_{i'\in\partial^-i}\bra{\sum_{i''\in\partial i^-\backslash i'}
\epsilon^{i''\to i} +\sum_{a\in\partial ^\Delta i}\epsilon^{a\to i}}\nonumber\\
	&=p\sum_{i'\in\partial^-i}\bra{\sum_{i''\in\partial i^-}
\epsilon^{i''\to i} +\sum_{a\in\partial ^\Delta i}\epsilon^{a\to i}-\epsilon^{i'\to i}}\nonumber\\
&=p\brb{(d_i^--1)\ein^{i-}+d_i^-\ein^{i\Delta}}.
\end{align}
Similarly, sum of messages coming from a triangle to node $i$ can be evaluated as
\begin{align}
	\lambda\ein^{i\Delta}&=\lambda\sum_{a\in\partial ^\Delta i}\epsilon^{a\to i}\nonumber\\
	&=q\sum_{a\in\partial^\Delta i}\sum_{j\in a\backslash i}\bra{ 
	\sum_{j'\in\partial j^-\backslash a}\epsilon^{j'\to j}+\sum_{b\in\partial j^\Delta\backslash a}\epsilon^{b\to j}}\nonumber\\
	&=q\sum_{a\in\partial^\Delta i}\sum_{j\in a\backslash i}\bra{ 
	\sum_{j'\in\partial j^-\backslash a}\epsilon^{j'\to j}+\sum_{b\in\partial j^\Delta}\epsilon^{b\to j}-\epsilon^{a\to i}}\nonumber\\
	&=q\sum_{a\in\partial^\Delta i}\sum_{j\in a\backslash i}\bra{ 
	\ein^{j-}+\ein^{j\Delta}}-q\eout^{i\Delta},
	\end{align}
	where $j\in a$ means $j$ is one end point of triangle $a$.
Sum of outgoing message going to triangles from node $i$ are written as
	\begin{align}
	\lambda\eout^{i\Delta}&=\lambda\sum_{a\in\partial ^\Delta i}\sum_{j\in a\backslash i}\epsilon^{a\to j}\nonumber\\
&=q\sum_{a\in\partial ^\Delta i}\sum_{j\in a\backslash i}
	\sum_{k\in a\backslash i,j}\bra{\ein^{k-}+\ein^{k\Delta}-\epsilon^{a\to k}}\nonumber\\
	&= 2q\bra{d_i^\Delta\ein^{i-}+d_i^\Delta\ein^{i\Delta}-\ein^{i\Delta}}+
	q\sum_{a\in\partial^\Delta i}\sum_{j\in a\backslash i}\bra{\ein^{j-}+\ein^{j\Delta}-\epsilon^{a\to i}}\nonumber\\
	&= 2q\brb{d_i^\Delta\ein^{i-}+(d_i^\Delta-1)\ein^{i\Delta}}+
	q\sum_{a\in\partial^\Delta i}\sum_{j\in a\backslash i}\bra{\ein^{j-}+\ein^{j\Delta}}-q\eout^{i\Delta}
\end{align}
Thus if we let $\ein^-=\{\ein^{i-},i\in\{1,...,n\}\}$,
$\ein^\Delta=\{\ein^{i\Delta},i\in\{1,...,n\}\}$,
$\eout^-=\{\eout^{i-},i\in\{1,...,n\}\}$,
$\eout^\Delta=\{\eout^{i\Delta},i\in\{1,...,n\}\}$,
then in a matrix form, above equations can be written as
\begin{align}
	\lambda
\begin{pmatrix}
	\ein^{-}\\
	\eout^{-}\\
	\ein^{\Delta}\\
	\eout^{\Delta}
\end{pmatrix} \, = C'\begin{pmatrix}
	\ein^{-}\\
	\eout^{-}\\
	\ein^{\Delta}\\
	\eout^{\Delta}
\end{pmatrix} \, 
\end{align}
with
\begin{align}
C' = \begin{pmatrix}
p \A^- & -p \id & p \A^- & 0 \\
p (\D^- - \id) & 0 & p \D^- & 0 \\
q \A^\Delta & 0 & q \A^\Delta & -q \id \\
2q \D^\Delta+q \A^\Delta & 0 & 2q(\D^\Delta-\id)+q \A^\Delta & -q \id
\end{pmatrix} \, . 
\end{align}
where $\id$ denotes the $n$-dimensional identity matrix, $\A^-$ is
the adjacency matrix of nodes connected by single edges; $\D^-$ is
the diagonal matrix of single-edge-degrees;$\A^\Delta$ is 
the adjacency matrix of nodes connected by triangles; $\D^\Delta$
is the diagonal matrix of triangle-degrees. That is
\begin{align}
	\A^-_{i,j} &= \begin{cases} \mbox{$1$, if $(i,j)\in\mathbb{E}$} \\ \mbox{$0$, other wise}  \end{cases}& 
%		\D^-_{i,j} &= \begin{cases} \mbox{$\sum_k\A^-_{ik}$ if $i=j$} \\ \mbox{$0$ other wise}  \end{cases} \\
\A^\Delta_{i,j} &= \begin{cases} \mbox{$1$, if $\exists (i,j,l)\in\mathbb{T}$ }\\ \mbox{$0$, other wise}  \end{cases} \\
%		\D^\Delta_{i,j} &= \begin{cases} \mbox{$\frac{1}{2}\sum_k\A^\Delta_{ik}$ if $i=j$} \\ \mbox{$0$ other wise}  \end{cases} \\
		\D^-_{i,i} &= d^-_i=\sum_k\A^-_{ik}&
		\D^\Delta_{i,i} &= d^\Delta_i=\frac{1}{2}\sum_k\A^\Delta_{ik}.
\end{align}

Thus we can see that $\lambda$ is both an eigenvalue of matrix $C$ and matrix $C'$. 
\hfill\ensuremath{\square}

%\section{Descriptions of networks listed in Table 1 of main text}
%Descriptions and citations are listed in table \ref{tab:1}.
%\begin{table}[t]
%  \centering
%\begin{tabular}{c| l}
%	\hline
%	\hline
%	Karate club \cite{Zachary1977}&Social network of members of a karate club.\\
%	Les Miserables \cite{Knuth93}&Network of characters in the novel \textit{Les Miserables}.\\
%	Political books\cite{politicalbooks} &Network of books of US politics.\\
%	Political Blogs \cite{Adamic2005}&Network of blogs on US politics in 2004.\\
%	Coauthorships  \cite{Newman2006}&Network of coauthorships of scientists working on network theory.\\ 
%	Protein  \cite{jeong2001lethality}&Network of protein interactions.\\
%	Power grid  \cite{Watts1998}&Network of topology of the Western State Power Grid of US.\\
%	Collaboration \cite{newman2001structure}&Network of coauthorships between high-energy physicists.\\
%	Email \cite{Leskovec2009}&Network of Enron email communications.\\
%	Gnutella \cite{matei2002mapping}&Gnutella pear-to-pear network.\\
%	SBM\cite{holland1983stochastic}&A network generated by the stochastic block model.\\
%	TSBM [unpublished]&A network generated by the stochastic block model with triangles.\\
%	\hline
%  \end{tabular}
%  \caption{\label{tab:1}}
%  \end{table}

%\section*{References}

%\medskip

%\clearpage
%\small
%\bibliographystyle{unsrtnat}
%\bibliographystyle{abbrvnat}
%\bibliographystyle{abbrv}
%\bibliography{/Users/pan/projects/ref/zp}

\end{document}